\documentclass[%
reprint,
superscriptaddress,
showpacs,
 amsmath,amssymb,
prl,
]{revtex4-1}

\usepackage{graphicx}% Include figure files
\usepackage{dcolumn}% Align table columns on decimal point
\usepackage{natbib, textcase,bm}

\begin{document}

%\preprint{}
\preprint{RAL Space}

\title{Mini-magnetospheres above the lunar surface and the formation of lunar swirls
}

\author{R.A. Bamford}
        \email{Ruth.Bamford@stfc.ac.uk} 
        \altaffiliation[Also at ]{York Plasma Inst, Dept of Physics, Uni. of York,
Heslington, York, YO10 5DQ, UK.}
\author{B. Kellett}
\author{W. J. Bradford}
	 \affiliation{RAL Space, Rutherford Appleton Laboratory, Chilton, Didcot, OX11 0QX,
UK.}
\author{C. Norberg}
        \affiliation{Swedish Institute of Space Physics, Box 812, SE-981 28 Kiruna, Sweden}
\author{A.Thornton}
\author{K. J. Gibson} 
        \affiliation{York Plasma Inst, Dept of Physics, Uni. of York, Heslington, York,
YO10 5DQ, UK.}
\author{I.A. Crawford} 
            \affiliation{Dept of Earth and Planetary Sciences, Birkbeck College, London}
\author{L. Silva} 
\author{L. Gargat\text{\'e}}
        \affiliation{Instituto Superior T\text{\'e}cnico, 1049-00, Lisboa, Portugal}
\author{R. Bingham}
        \altaffiliation[Also at ]{Central Laser Facility, Rutherford Appleton Laboratory, Chilton, Didcot, OX11 0QX, UK.}
        \affiliation{University of Strathclyde, Glasgow, Scotland, UK.}

\date{\today}% It is always \today, today,

\begin{abstract}

In this paper we present in-situ satellite data, theory and laboratory validation that show how small scale collisionless shocks and mini-magnetospheres can form on the electron inertial scale length. The resulting retardation and deflection of the solar wind ions could be responsible for the unusual ``lunar swirl'' patterns seen on the surface of the Moon.

\end{abstract}

%\pacs{96.20.Jz, 52.35.Tc, 52.72.+v,94.05.-a, 94.30.vf, 96.20.Dt,96.12.jk}

% 96.20.-n	Moon
%96.20.Jz	Gravitational field, selenodesy, and magnetic fields
%52.35.Tc	Shock waves and discontinuities
%52.72.+v	Laboratory studies of space- and astrophysical-plasma processes 
%94.05.-a	Space plasma physics 
%94.30.vf	Solar wind/magnetosphere interaction
%96.20.Dt	Features, landmarks, mineralogy, and petrology
%96.12.jk	Magnetospheres

%\keywords{}
%\keywords{mini-magnetospheres, Moon, magnetic anomalies, lunar swirls, collisionless shocks, Larmor orbit}
%Use showkeys class option if keyword

\maketitle

Miniature magnetospheres have been found to exist above the lunar surface \cite{Lin98} and are closely related to features known as ``lunar swirls" \cite{Kramer2011}. Mini-magnetospheres exhibit features that are characteristic of normal planetary magnetospheres namely a collisionless shock. Here we show that it is the electric field associated with the small scale collisionless shock that is responsible for deflecting the incoming solar wind around the mini-magnetosphere. These ions impacting the lunar surface resulting in changes to the appearance of the albedo of the lunar ``soil''\cite{Kramer2011}. The form of these swirl patterns therefore, must be dictated by the shapes of the collisionless shock. 

Collisionless shocks are a classic phenomena in plasma physics, ubiquitous in many space and astrophysical scenarios \cite{Sagdeev1967}. Well known examples of collisionless shocks exist in the heliosphere, where the shock is formed by the solar wind interacting with a magnetised planet. What is a surprise is the size of the mini-magnetospheres, of the order of several 100~km; orders of magnitude smaller than the planetary versions. Results from various lunar survey missions have built up a good picture of these collisionless shocks.

These collisionless shocks have a characteristic structure in which the ions are reflected from a rather narrow layer, of the order of the electron skin depth $c/\omega_{pe}$ (where $c$ is the speed of light and $\omega_{pe}$ is the electron plasma frequency), by an electrostatic field that is a consequence of the magnetised electrons and unmagnetised ions. The narrow discontinuity in the shock structure produces a specular reflected ion component with a velocity equal to or greater than the incoming solar wind velocity. The reflected ions from a counter-propagating component to the solar wind flow that form the magnetic foot region, which extends about an ion Larmor orbit upstream from the shock. This occurs when the Mach number (the ratio of flow velocity to Alfv\'en velocity) is of the order 3 or less. 

We have carried out laboratory experiments using a plasma wind tunnel, to investigate mini-magnetospheres and found that they show characteristics similar to the lunar mini-magnetospheres. A quantified comparison between the observations, both in space and in the laboratory, with theoretical values shows excellent agreement.
\begin{figure}
       \includegraphics[width=0.5\textwidth]{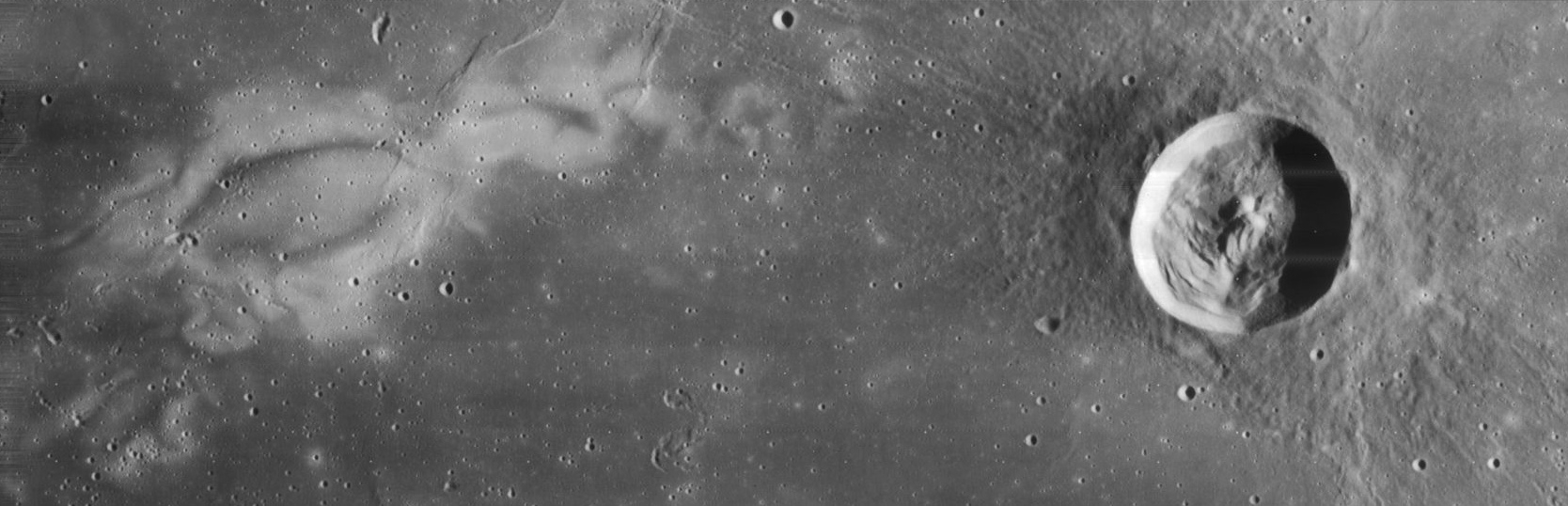} 
       \caption{The Reiner Gamma formation (7.4$^\circ$N, 300.9$^\circ$E) is an example of a lunar swirl. Pictured here on the left hand side of the image. Reiner Gamma is named after the Reiner impact crater shown for comparison on the right. The crater is 117km to the east and has diameter of 30km with a depth of 2.6km. By contrast the unusual diffuse swirling of the formation and concentric oval shape has fluid-like ``wisps'' that extend further to the east and west. Its distinctive lighter colour stands out against the flat, dark surface of Mare Oceanus Procellarum. Unlike crater ejecta, the tilted``$\gamma$'' shape of the formation appears unrelated to any topographic structures that would account for its presence. Image courtesy of NASA.}
        \label{FIG:swirlphoto}
\end{figure}

The Reiner Gamma Formation shown in Figure~\ref{FIG:swirlphoto}, is one example of a number of small ``swirls'' of apparently lighter coloured material visible on the lunar surface. These distinctive patterns do not appear to correlate with other surface features - such as impact craters or mountains and valleys, but do coincide with patches of significant magnetic field \cite{Kramer2011}. All of these lunar swirls have been found to be associated with magnetic anomalies \cite{Kramer2011}. The appearance of lighter albedo material on the Moon is usually indicative of the presence of younger or less weathered lunar material \cite{McKay1991}. Explanations that account for these changes in the albedo are either a) that the lighter regions have been ``shielded'' from receiving the same solar wind flux as the surrounding regolith and hence appear younger \cite{Kramer2011}, or b) that the lighter colour material originates from just below the surface and has been lifted up and deposited on the top of older{/}darker regolith \cite{GarrickBethell2011}.  The sharpness of the lunar swirl formations is enhanced by the contrast of ``dark lanes'' (suggesting locally enhanced solar wind proton bombardment) within the high--albedo swirls.

\begin{figure}
        \includegraphics[width=0.45\textwidth]{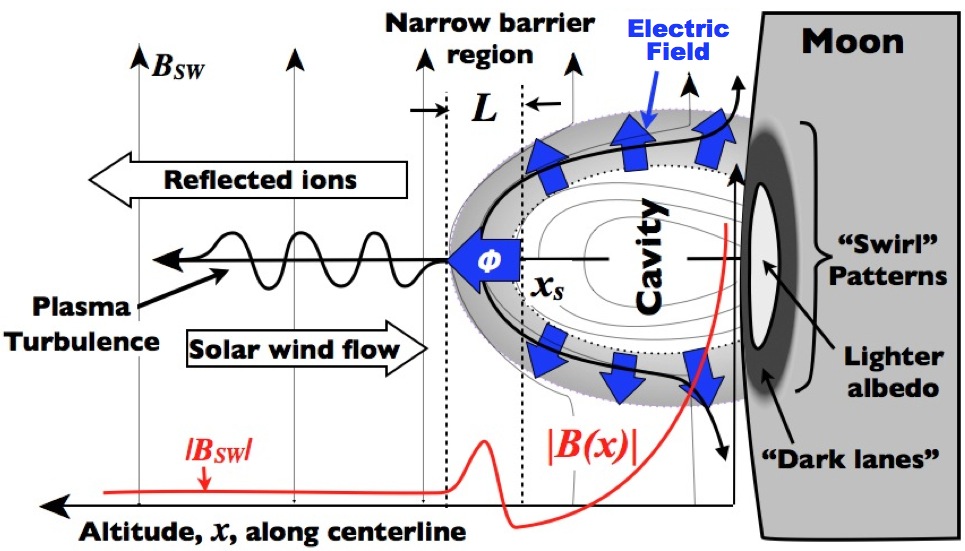}
        \caption{ A sketch of the generic scenario of a miniature  magnetic field emerging from the lunar surface and interacting with the solar wind. Flow is from left to right.} \label{FIG:Alfvenlikesketch}
\end{figure}

Evidence by in-situ measurements from  space craft including  Lunar Prospector (1998-1999) \cite{LP}, Kaguya (2007-2009) \cite{KAGUYA}, Chandrayaan--1 (2008-2009) \cite{Chandrayaan} and Nozomi (1998) spacecraft \cite{Nozomi}, are consistent with the presence of collisionless shocks and the formation of mini-magnetospheres. Because the observational data derives from a sequence of case studies from different missions\cite{Lin98, Halekas2008a, Wieser2010, Hashimoto2010, Saito2008, Saito2010, Futaana2010, Futaana2003}, it is to be expected that there is some variation in consistency. However a schematic picture of the interaction can be unravelled from the specific observations to form a simplified, generic model. This is illustrated in Figure~\ref{FIG:Alfvenlikesketch}.

A region of enhanced magnetic field strength (by factors of 2 to 3) is observed at an altitude $x_s$, above the lunar surface where the magnetic intensity from the magnetic anomaly reaches pressure balance with the plasma pressure from the solar wind. The magnetic field components were observed to rotate in a fashion consistent with the spacecraft passing through a region in which the solar magnetic field was being ``draped'' around a small magnetic obstacle or ``bubble'' \cite{Lin98}. Within the narrow barrier region is a low density cavity seen in the ion data \cite{Wieser2009}. The barrier region is of the order of kilometers across. Just ahead of a large ramp in magnetic field strength there is enhanced magnetic field turbulence in the solar wind magnetic field\cite{Lin98}. This is accompanied by electrostatic solitary waves at the lower hybrid frequency and electron fluxes abruptly increase and their energy distribution changes indicating that electrons are energized and not simply compressed \cite{Lin98, Halekas2008a, Hashimoto2010, Saito2008, Futaana2010, Wieser2010}. Intense electrostatic waves, of frequencies of the order of the plasma lower-hybrid frequencies (1-10Hz) were recorded over the locations of the magnetic anomalies by the Kaguya spacecraft \cite{Saito2008,Hashimoto2010}. Variations in the intensity of these waves over magnetic anomalies with solar wind pressure suggests a dynamic interaction \cite{Hashimoto2010}. At 100km Kaguya observed  protons reflected back from the magnetic structures with greater energies (by factors 3 to 6) than the incident solar wind flux \cite{Saito2008}. Chandrayaan--1 also observed back streaming protons accelerated by similar factors close to the shock surface \cite{Wieser2009, Wieser2010}. These higher energy protons are accelerated by the convective electric field seen by the reflected protons in the solar wind flow \cite{Saito2008}. Below 50km altitude, proton back scattering disappeared \cite{Saito2010}, indicating that the flowing solar wind ions were no longer reaching these altitudes, and suggesting that the spacecraft was flying through a solar wind plasma ``cavity''. Chandrayaan--1 provided two-dimensional maps of the spatial extent of the cavity above the magnetic  anomaly near the Gerasimovic crater \cite {Wieser2010}. The dimensions of the cavity and the magnetic field anomaly ($\sim$360km across) were very similar. The cavity was more distinct in the higher energy ions $>$150 eV than at lower energies. The overall size was about 360 km in diameter, coincident with the center of the magnetic anomaly. An outer ring of about 300 km wide of enhanced proton flux suggested that the incoming solar wind ions were being deflected around from the central ``bubble'' to impact the surface in this relatively narrow surrounding region. Observations provided by the Nozomi spacecraft in 1998, suggest that  characteristic wakes arising from the lunar mini-magnetospheres can extend to significant altitudes (2800 km). The instrumentation on-board recorded two peaks in density of non-thermal protons either side of a cavity of reduced ion flux \cite{Futaana2003}. These observation agree with those of Chandrayaan- 1 at intermediate altitudes above the mini-magnetosphere bow shock \cite{Futaana2010}.

The key to understanding how such structures can arise is to use a two fluid model of the plasma, in which the ions in the flowing plasma are un-magnetised and the electrons are magnetised. As the solar wind with its embedded magnetic field impacts a magnetic structure, a cavity is created. The cavity is bounded by an enhanced magnetic field that is about a factor three greater than the solar wind field with a width, $L$, estimated to be similar to the electron skin depth. This field traps a low density plasma that forms part of the barrier. The cavity is created by the induced currents that flow on its outer boundary, giving rise to magnetic field  enhancement that opposes the penetration of the solar wind  according to Lenz's law. The enhanced magnetic field drapes around the cavity.

The magnetic field enhancement shown in the satellite observations, controls the flow of  the electrons that are magnetized on these scales. The solar wind electrons are slowed and deflected by the magnetic structure. The ions, on the other hand, due to their greater inertia, cannot respond sufficiently promptly to abrupt or sudden changes in magnetic field on these scales. These unmagnetized ions can therefore easily penetrate through the barrier. Ions flowing through the barrier result in a space charge separation between the electrons and ions, forming an electric field that is responsible for  slowing and  reflecting/deflecting of them over distances of the order of the electron skin depth \cite{Phelps1973}.

The equation of motion that controls the behaviour of the ions is \cite{Woods1987} :
\begin{equation}
        n_{i}\frac{d}{dx}(\frac{1}{2}m_{i}v^{2}_{i}+e\phi)+\frac{dp_{i}}{dx}=0
\end{equation}
where $\phi$ is the electric potential ramp across the barrier and $p_i$ is the ion pressure $=n_{i}kT_{i}$ and $v_{i}$ is the ion velocity, the bulk value of which is the solar wind speed, $\sim v_{sw}$.

The expression for the electric potential component, $\phi$, responsible for slowing and deflecting the ions is \cite{TidmanKrall1971, Bingham1993}:
\begin{equation}
          \phi =  - \frac{1}{2\mu_0n e}.B_z^2
\end{equation}
If the instantaneous density, $n$, here is $\sim 5 \times 10^{6}m^{-3}$ and a magnetic field $B_z \sim30 \times 10^{-9}T$ values similar to those  observed at the pile-up reported by Lunar Prospector \cite{Lin98}, then the mean value from (2) of the potential would be $\phi_{theory} \sim 450 V$. This average value is very similar to the $\phi_{obs} \sim 400V$  \cite{Futaana2010} and would account for the observed counter-streaming protons.  Although the force does act initially on the electrons, the resulting electric field formed then acts on the ions. The consequence is this force acts to keep particles out of regions of high magnetic field. The shock thickness of the electric field, $L$, is much narrower than the ion Larmor radius. The ions experience a sufficient impulsive force in a direction normal to the barrier to reverse their velocity. These ions move upstream with a velocity $2v_{sw}$ in the solar wind frame and form a broad much thicker region than the barrier sometimes known as the shock foot.  Within the shock foot region, the counter streaming ions are responsible for a number of micro-instabilities, such as the modified two stream instability \cite{McBride1972}, that drives plasma wave turbulence close to the lower hybrid frequency $ = \left ( \omega_{ce}\omega_{ci} \right )^{1/2}$, where $\omega_{ce}$ and $\omega_{ci}$ are the electron and ion cyclotron frequencies respectfully. This would agree with data from all the spacecraft that observed intense lower hybrid electrostatic oscillations of the order of 1-10 Hz \cite{Futaana2003}. The reflected or deflected ions can also form a non-thermal ring distribution in ion velocity space due to $E \times B$ pick--up exactly as reported \cite{Futaana2003}. Alternatively a simple counter-streaming population would be observed depending upon the particular conditions intersected by the spacecraft on its fly through. Both are consistent. The changes in the particle distributions observed by the in-situ spacecraft, between the ions and electrons either side of the shock result from the formation and interaction of lower-hybrid waves generated close to the bow shock \cite{BinghamBamford2010}. The resonant interaction between lower-hybrid turbulence and electrons can result in field aligned electron acceleration \cite{BinghamBamford2010}. These waves are most probably excited by the modified two stream instability driven by reflected ions. 

The density of transformed electrons, $n_{Te}$, is estimated by balancing the growth rate of the instability initiated by pickup ions with Landau damping due to electrons moving parallel to the magnetic field. Estimations for the average energy $\epsilon_e$ of the accelerated electrons and their number density $n_{Te}$ can be made\cite{BinghamBamford2010}: 
\begin{equation}
        \epsilon_e \approx \alpha^{2/5} \left ( \frac{m_e}{m_i} \right )^{1/5} m_i v_{sw}^2
\end{equation}
\begin{equation}
        n_{Te} \approx n_i \alpha^{2/5} \left ( \frac{m_e}{m_i}\right )^{1/5}
\end{equation}
where $\alpha$ is the energy transformation efficiency from ions to electrons (with masses $m_i$ and $m_e$ respectfully). 

\begin{table}
\caption{A comparison of the absolute and dimensionless distances (normalized to $c/\omega_{pi}$), in space and the laboratory experiment. \label{TABLE:lab}}
        \begin{ruledtabular}
        \begin{tabular}{l | c  c || c  c }
 	                           & \multicolumn{2}{c}{\textrm{Space}} & \multicolumn{2 }{c}{\textrm{Laboratory}}\\
\textrm{Parameter in SI Units}   	& \textrm{Value}  &\textrm{Dim{'}}&\textrm{Value}&\textrm{Dim{'}}\\
\hline
\textrm{Thermal energy in eV} 	         &     5   		&                     &        5                &                         \\
\textrm{Density},$m^{-3}$                 &    $5.10^6$     	&                     &    $10^{17}$      &                         \\
\textrm{Flow speed, $ms^{-1}$}        &  $4.10^{5}$     &                     &    $8.10^{4}$    &                         \\
\textrm{Magnetic field strength, $T$}& $10^{-8}$	       &                     &     0.03              &                        \\
\textrm{Plasma  Beta}      			  &         0.1   	&                     &     $2.10^{-4}$  &                         \\
\textrm{MACH\# Acoustic,Alfv\'enic } &         20,  5  	&                     &     3.5, 0.03      &                         \\
\textrm{Flow Mean free path}, $m$   &  $10^{16}$ 	&  $10^{11}$   &         300          &       400           \\
\textrm{Debye length, $m$}               &         7  		&  $10^{-5}$   &    $5.10^{-5}$   &    $7.10^{-5}$  \\
\textrm{Electron Larmor radius, $m$}&        800  	& $8.10^{-3}$  &   $3.10^{-4}$  &   $4.10^{-4}$   \\
\textrm{Thermal Ion Larmor radius}  &  $6.10^{4}$	&         0.6       &    $10^{-2}$     &       0.01           \\
\textrm{Flow Ion Larmor radius, $m$}&  $5.10^{5}$ 	&          5         &    $3.10^{-2}$   &      0.04            \\
\textrm{Electron Skin depth, $m$}      &    $2.10^{3}$ &      0.02         &     $2.10^{-3}$ &       0.01            \\
\textrm{Ion Skin depth $c/\omega_{pi}$, $m$}
                                                           &     $10^{5}$    &        1            &         0.7          &         1              \\
               \end{tabular}
        \end{ruledtabular}
\end{table}

The energy transformation coefficient $\alpha$, has only a weak influence on the result \cite{BinghamBamford2010}. Therefore using a value of $\alpha \sim 0.1$, the case near a bow shock, with an ion energy of $ \sim 1 keV$  and electron energy $\epsilon_e \sim 100 eV$, the density of the accelerated electrons would be 10\% of the ion density ($n_{Te}  \simeq 0.1n_i $). This is consistent with the typical values reported by Kaguya \cite{Saito2010} and Chandrayaan--1 \cite{Lue2010}. This simple estimate demonstrates the large efficiency of lower-hybrid waves as an acceleration mechanism for the electrons. The result is a more efficient boundary than would be predicted by MHD (magnetohydrodynamics); comparisons with the particle distribution data could confirm this. The theoretical width of the barrier is expected to be of the order of the electron skin depth, $L=$1 to 2 km, and not the 100's km of the ion skin depth \cite{Phelps1973}. See Table~\ref{TABLE:lab} for typical values.

Computational simulations of kinetic processes are non-trivial. A scaled laboratory experiment has the potential to deliver physical insights and to offer observational signatures which can be examined, to establish whether they are consistent or not with the in-situ space data. The Plasma Wind Tunnel \cite{Rusbridge2000} used here shares the same phenomenological regime as the subset of analogous space parameters. These are: the plasma is collisionless, the bulk flow speed is supersonic and the electrons are magnetized but the ions are not. The absolute and dimensionless parameter are given in Table~\ref{TABLE:lab}. 
\begin{figure}
       \includegraphics[width=0.48\textwidth]{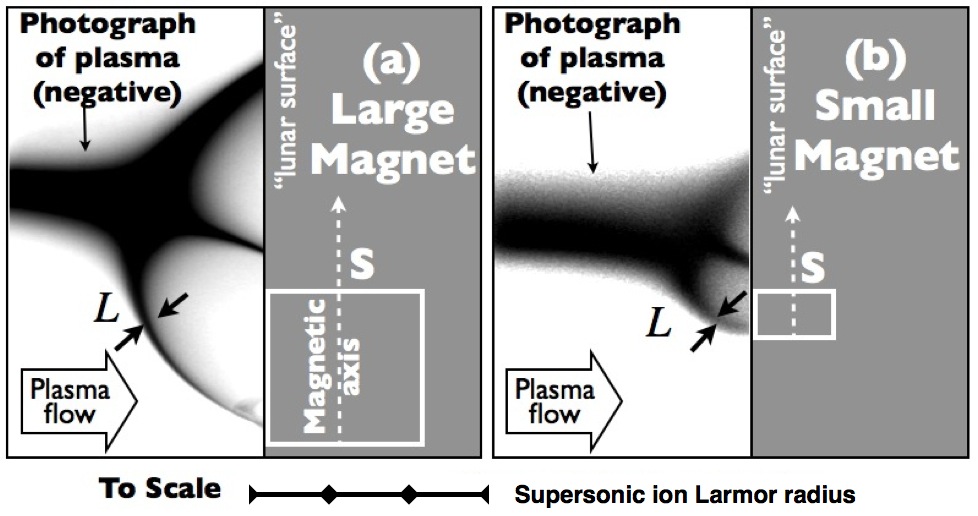} 
        \caption{Photographs, with graphical labels overlaid, of the supersonic plasma stream being deflected by two different strength magnets. (a) 23x20mm, 0.45 T on axis. (b)10x3mm, 0.30T on axis.  External horizontal magnetic field is 0.03T. A field-aligned current can be seen connecting the `cusp' region to the southern poles in both cases. } \label{FIG:MiniMaglab_photos}
\end{figure}
\begin{figure}
        \includegraphics[width=0.48\textwidth]{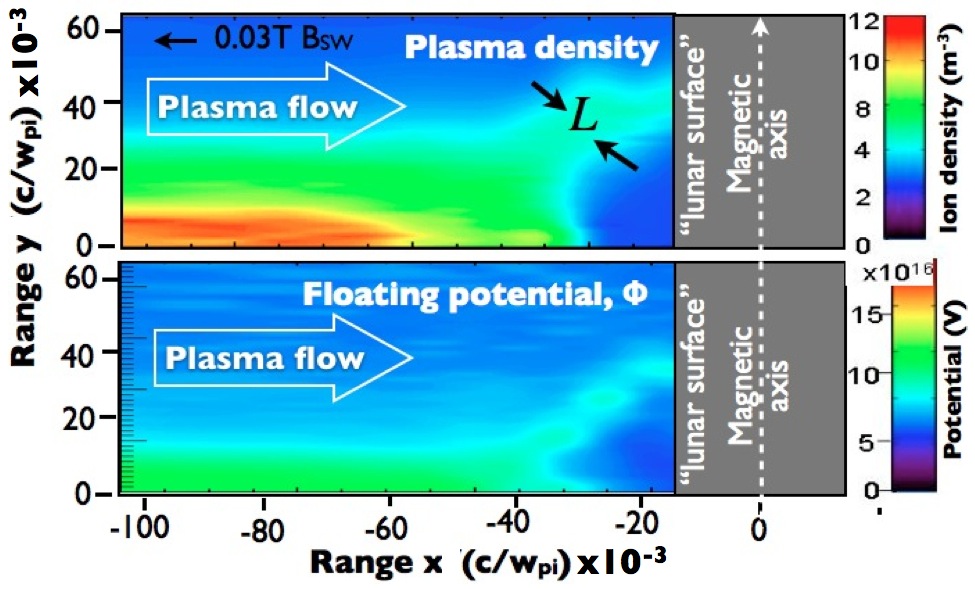}
        \caption{ In-situ Langmuir probe plots of the plasma density and floating potential ahead of the larger target magnet (23x20mm, 0.45T). Data from \cite{Bamford2009}.} \label{FIG:MiniMaglab_ProbeData1}
\end{figure}
\begin{figure}
	\includegraphics[width=0.45\textwidth]{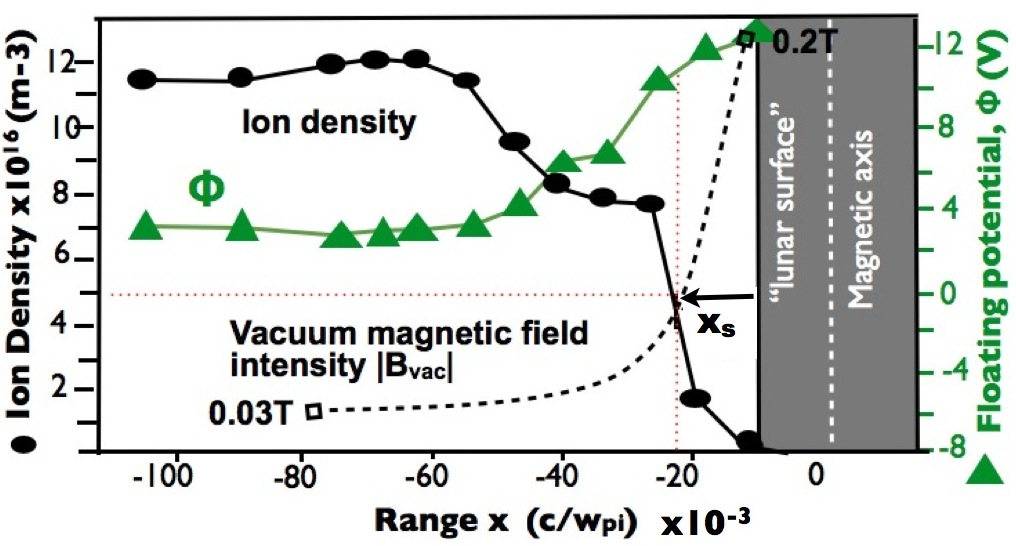}
        \caption{A plot of the measured ion density (black dots) and electric potential (green triangles) along the axis of the experiment in the vertical plane. The plasma flow is from left to right. Data from\cite{Bamford2009}.}
\label{FIG:MiniMaglab_ProbeData2}
\end{figure}

In Figure~\ref{FIG:MiniMaglab_photos} we show two photographs of the Plasma Wind Tunnel in action. The supersonic hydrogen plasma stream is here impacting the magnetic dipole fields of two different sized magnets under the same conditions. The plasma stream appears dark on the photograph and is enhanced for clarity. The structures are 3D so some blurring due to line-of-sight occurs. Despite the $>\times2$ difference in the size of the magnetic ``bubble'' obstacle in the two cases, the plasma stream is corralled the same way into a narrow boundary layer and deflected around the outsides to form a cavity. The key inset shows the ion Larmor radius to scale with the photographic data demonstrating that the interaction is on the sub-ion Larmor radius scale as with the Lunar case.

The Langmuir probe data shown in Figure~\ref{FIG:MiniMaglab_ProbeData1} corresponds to the larger of the two magnets shown in Figure~\ref{FIG:MiniMaglab_photos}. The ion density shown in the upper panel confirms the formation of a cavity or void in the plasma stream. The floating electric potential (lower panel) coincides with the locations of the measured ion density shown in the upper panel. This demonstrates that the ions are indeed electrostatically confined on this scale. The measured width of the boundary layer is between 2 to 3mm, similar to the calculated electron skin depth of $\sim$2mm \cite{Phelps1973} and is very much less than the calculated ion inertial length of 700mm or the 30mm of the ion Larmor radis (see Table~\ref{TABLE:lab}). The quantified relation between the ion density and the electric potential can be better compared in the 1D plot shown in Figure~\ref{FIG:MiniMaglab_ProbeData2}. At the point $x_s$ shown in the figure, the measured electric potential is $\phi_{obs} \sim 8V$. Here again, the estimate provided by Equation (2) provides a good match to the observed value with $\phi_{theory} \sim 12.5V$ (using $B=0.05T$, $n=5\times10^{16}m^{-3}$). This shows that the dimensionless analysis holds on a very different absolute scale to the case in space.

The analysis presented here shows that the strength of the deflecting electric field of a mini-magnetosphere and collisionless shock, is not dependent on  the overall size of the magnetic ``bubble'' but is related to the local \textit{gradient} in the magnetic field strength. None of the real features are simple single dipoles. Close to the surface, the magnetic topology in a magnetic anomaly is likely to be very irregular, comprising a range of overlapping cavities and gradients. This would lead to a pattern of retarded and accelerated space weathering and hence areas of lighter material with embedded ``dark lanes''. A further range of contrasts in the surface ``aging'' would come from the variations in the impacting plasma wind environment with solar activity, the Moon's orbit in and out of the Earth's magnetosphere and lunar phases.

In conclusion, the model of small scale collisionless shocks that we present agrees with a laboratory plasma wind tunnel experiment confirming the presence of a narrow electrostatic potential of thickness of the order of the local electron skin depth, and that it is the force responsible for the control and deflection of the ions. All the observational data from spacecraft is quantifiably consistent with the theoretical model. 

\section{Acknowledgements}
The authors would like to thank Science and Technology Facilities Research Council's Center for Fundamental Physics for support.

%\bibliographystyle{unsrt} % unsrt - numbered according to appearance
%\bibliography{MiniMagospheres_on_Moon_PRL}

%\section{Figures}

\bibliography{Bamford_PRL_text_figs}

\end{document}